\documentclass[aps,pra,twocolumn,showpacs,floatfix,groupedaddress]{revtex4}

\usepackage{graphicx}
\usepackage{dcolumn}
\usepackage{bm}

\begin{document}

\title{Bose-Einstein condensation of trapped atoms with dipole interactions}

\author{Kwangsik Nho and D. P. Landau}

\affiliation{Center for Simulational Physics, University of Georgia, Athens,
Georgia 30602}

\date{\today}

\begin{abstract}
The path integral Monte Carlo method is used to simulate dilute
trapped Bose gases and to investigate the equilibrium properties 
at finite temperatures. The quantum particles have a long-range dipole-dipole
interaction and a short-range $s$-wave interaction. 
Using an anisotropic pseudopotential for the long-range dipolar interaction 
and a hard-sphere potential for the short-range $s$-wave interaction, we
calculate the energetics and structural properties 
as a function of temperature and the number of particles. 
Also, in order to determine the effects of 
dipole-dipole forces and the influence of the trapping field on the dipolar 
condensate, we use two cylindrically symmetric harmonic confinements 
(a cigar-shaped trap and a disk-shaped trap). We find that the net effect 
of dipole-dipole interactions is governed by the trapping geometry. For a
cigar-shaped trap, the net contribution of dipolar interactions is attractive 
and the shrinking of the density profiles is observed. For a disk-shaped trap, 
the net effect of long-range dipolar forces is repulsive and 
the density profiles expand.
\end{abstract}

\pacs{03.75.Hh, 03.75.Nt, 05.30.Jp, 02.70.Ss}

\maketitle

\section{Introduction}
The experimental observation of Bose-Einstein condensation (BEC) 
in dilute trapped and supercooled atomic vapors~\cite{ADB} has stimulated
enormous experimental and theoretical interest in the physics of
the weakly interacting Bose gas (see \cite{DGPS} for a review). 
Especially intriguing is the role of 
dimensionality and two-body interactions. Lower dimensional atomic gases have 
been realized with strong quantum confinement in one or more directions.
These systems exhibit very peculiar properties, such as the exact mapping
between interacting bosons and noninteracting fermions 
(Tonks-Girardeau gas)~\cite{TG}, 
because the role of quantum fluctuations and strong correlations is enhanced. 
Although these gases are very dilute, BEC of trapped atomic gases
is strongly influenced by two-body interactions. Within a mean-field
approximation, the strength of the interatomic interactions can be typically 
characterized by a single parameter, the $s$-wave scattering length $a_{s}$,
at very low temperatures. This van der Waals type interaction is isotropic 
and short-range in comparison to the average interparticle separation.
Accordingly, 
Bose-Einstein condensates created thus far have been very accurately
described by the Gross-Pitaevskii theory (see \cite{DGPS} for a review), 
a mean-field approach. 
Furthermore, it has been possible to tune 
the s-wave scattering length $a_{s}$ to essentially any value, 
positive or negative, 
by utilizing a magnetic atom-atom Feshbach resonance~\cite{FR}.

Studies of the influence of the interparticle interactions 
on the properties of BEC have focused
mainly on the short-range interactions~\cite{DGPS}. 
However, understanding new kinds of systems will require the investigation
of the effects of additional interactions,
since Bose-Einstein condensation of molecules has now been observed 
using new techniques for creating ultracold molecules~\cite{GRJ}. 
In addition,
Ling {\it et al.}~\cite{Ling} proposed an efficient method 
to convert atomic condensates 
into stable molecules using a generalized Raman adiabatic passage scheme.
Very recently, the generation of a BEC 
in a gas of $^{52}$Cr atoms was reported~\cite{GWHSP}.
As molecules can potentially have large dipole moments and a chromium atom has 
a very high magnetic moment of 6$\mu_{B}$ (Bohr magneton) in its ground state, 
BEC with dipole-dipole interactions in ultracold 
gases has attracted considerable theoretical attention over the past few 
years~\cite{Theory,YY1,Der,SSZL,GSL,GOK,SSL2,OGK2,OGE,DD,GGP}. 
Compared to the short-range interaction, 
the dipole-dipole interactions are long-range, anisotropic, and 
partially attractive. As a result, nonlinear effects due to long-range 
dipole-dipole interactions lead to new interesting properties;
different quantum phases such as 
superfluid, supersolid, Mott insulator, checkerboard, 
and collapse phases~\cite{GSL}, a self-bound Bose condensate in the field of 
a traveling wave~\cite{GOK}, a roton minimum 
in the excitation spectrum~\cite{SSL2,OGK2}, and promising candidates
for the implementation of fast and robust quantum-computing 
schemes~\cite{DD}.
First of all, Yi and You~\cite{YY1} suggested an effective pseudopotential for 
anisotropic dipolar interactions in the Born approximation and recently
Derevianko~\cite{Der} derived a more 
rigorous velocity-dependent pseudopotential 
for dipole-dipole interactions.  
Santos {\it et al.}~\cite{SSZL} 
showed a strong dependence of the stability of trapped dipolar BEC on the 
trapping geometry.
O'Dell {\it et al.}~\cite{OGE} derived exact results 
in the Thomas-Fermi regime for the statics and dynamics of a trapped 
dipolar BEC.
However, in all the theoretical studies so far the T=0 K situation has been 
investigated. In this paper we address the properties of trapped condensates 
with dipolar interactions at finite temperatures.
Several sources for dipolar BEC are atoms polarized by 
an electric field or atoms with large magnetic moments 
in addition to polar molecules.
The magnitude and sign of dipole-dipole interactions can be changed by rapidly
rotating an external field~\cite{GGP}. 
Thus, by varying the strength of the dipolar 
interactions relative to the s-wave scattering 
(tuning the dipolar interaction or changing 
the s-wave scattering length $a_{s}$) it is possible to study the effects 
of the dipolar interaction on a condensate in detail.

In this paper we attempt to understand Bose-Einstein condensation of
trapped dipolar gases, where the interparticle interaction consists of 
a long-range dipole-dipole interaction in addition to the usual short-range 
s-wave interaction (a purely repulsive hard-sphere potential). 
In particular, we demonstrate the influence of 
dipole-dipole forces on the properties of ultracold gases by comparing 
the energetics and structural properties of harmonically trapped atomic 
condensates with dipolar interactions with those for which only short-range 
interactions are present. In addition, we study the dependence of the sign 
and the value of the dipole-dipole interaction energy on the trapping geometry
by using two different types of cylindrical traps ( a cigar-shaped trap and
a disk-shaped trap). 
For these purposes, we use a finite-temperature path-integral 
Monte Carlo (PIMC) method~\cite{Ceperley}.

Our paper is structured as follow. Section~\ref{sec2} introduces 
the Hamiltonian of trapped dipolar hard-sphere systems used in this studies 
and an effective pseudopotential to describe anisotropic dipolar interactions 
in the Born approximation.
The path integral Monte Carlo technique briefly is also reviewed 
in Sec.~\ref{sec2}.
Section~\ref{sec3} presents our simulation results.
Finally, we conclude in Sec.~\ref{sec4}.

\section{Description of the system and Simulation Method}
\label{sec2}
We wish to study the problem of a quantum $N$-particle system to consider 
a harmonically trapped Bose-Einstein condensation in three dimensions. 
The quantum particles have both a long-range dipole-dipole interaction 
and the usual short-range hard-sphere interaction.
We assume that all atomic dipole moments are equal and oriented
along the $z$-axis (trap axis). Within the Born approximation, 
the dipole-dipole interaction 
potential between two dipoles separated by ${\bf r}_{ij}$ is given by 
\begin{eqnarray}
v^{d}(r_{ij}) = \frac{\mu^{2}(1-3\cos^{2}\theta_{ij})}{r_{ij}^{3}},
\end{eqnarray}
where $\mu$ characterizes the dipole moment, 
${\bf r}_{ij} = {\bf r}_{i} - {\bf r}_{j}$ is the vector between
the dipoles $i$ and $j$, and $\theta_{ij}$ the angle between the vector 
${\bf r}_{ij}$ and the direction of the dipoles (${\bf z}$). To date,
this pseudopotential has worked
reasonably well for the long-range dipole-dipole interaction. 
The corresponding Hamiltonian is then
\begin{eqnarray}
H &=& H_{0} + \sum_{i>j}^{N}v^{d}(r_{ij}) 
+ \sum_{i>j}^{N}v(r_{ij})\\
H_{0} &=& -\frac{\hbar^{2}}{2m}\sum_{i=1}^{N} \nabla_{i}^{2} + \frac{1}{2}m
\sum_{i=1}^{N}
(\omega^{2}_{x}x_{i}^{2}+\omega^{2}_{y}y_{i}^{2}+\omega^{2}_{z}z_{i}^{2})
\nonumber
\end{eqnarray}
where $H_{0}$ is the Hamiltonian for trapped ideal boson gases and $v(r)$
is the hard-sphere potential defined by 
\begin{eqnarray}
v(r) & = & +\infty \hspace{1in}(r < a_{s}) \nonumber\\
     & = &  0      \hspace{1.2in}(r > a_{s}).
\end{eqnarray}
In order to investigate the dependence of the sign 
and the value of the dipole-dipole interaction energy on the trapping geometry,
we consider Bose gases under a cylindrically harmonic confinement 
($\omega_{x} = \omega_{y} = \lambda \omega_{z}$ and $\omega_{z}$), 
where $\lambda$ is the aspect ratio.

Consider a system of $N$ dipolar hard spheres. The statistical mechanics of
a quantum system at an inverse temperature $\beta$, where 
$\beta = (k_{B}T)^{-1}$ and $k_{B}$ denotes the Boltzmann constant, is 
governed by the density matrix. The Bose-symmetrized density matrix is given by
\begin{eqnarray}
\rho_{B}({\bf \cal{R}},{\bf \cal{R}}';\beta) = \frac{1}{N!}\sum_{P}
\rho({\bf \cal{R}},P{\bf \cal{R}}';\beta),
\label{eq-dm}
\end{eqnarray}
where ${\bf \cal{R}}$ and ${\bf \cal{R}}'$ are two configurations 
of $N$ dipolar hard spheres, which denote $3N$ dimensional vectors, 
${\bf \cal{R}}$ = $(
{\bf R}_{1},\cdot\cdot\cdot,{\bf R}_{N})$.
$P$ denotes a permutation of particle labels among hard spheres and 
$P{\bf \cal{R}}$
is one such permutation. The inclusion of particle permutations
is crucial and superfluidity is an immediate consequence of Bose symmetry.
However, evaluating the density matrix for interacting systems
at very low temperatures is complicated by the fact that the kinetic and
potential terms in the exponent of the density matrix do not commute.
In order to avoid this problem, we insert $M-1$ intermediate configurations 
into Eq.~(\ref{eq-dm}):
\begin{eqnarray}
\rho({\bf \cal{R}},P{\bf \cal{R}}';\beta) & = & \int\cdot\cdot\cdot\int 
d{\bf \cal{R}}_{1}d{\bf \cal{R}}_{2} \cdot\cdot\cdot d{\bf \cal{R}}_{M-1} 
\nonumber \\
                & \times & \rho({\bf \cal{R}},{\bf \cal{R}}_{1};\tau)
\cdot\cdot\cdot\rho({\bf \cal{R}}_{M-1},P{\bf \cal{R}}';\tau),
\label{eq-dm2}
\end{eqnarray}
where $\tau$ = $\beta/M$ is the imaginary
time step and $M$ is referred to as the number of time slices.
Eq.~\ref{eq-dm2} is the path-integral formulation of the density matrix, 
which contains the discrete sum over permutations and multiple integrations 
over density matrices at a higher temperature $(k_{B}\tau)^{-1}$. 
Eq.~\ref{eq-dm2} can be evaluated in the path integral Monte Carlo technique
by a stochastic sampling of the discrete paths  $\{{\bf \cal{R}}, 
{\bf \cal{R}}_{1}, {\bf \cal{R}}_{2}, \cdot\cdot\cdot, {\bf \cal{R}}_{M-1}, 
P{\bf \cal{R}}'\}$ using
multilevel Monte Carlo sampling~\cite{Ceperley,Binder}, 
which accounts for permutations.
In order to use Monte Carlo sampling, we must first provide a pair-product 
form of the exact two-body density matrices. 
For the short-range hard-sphere potential,
we used the high-temperature approximation for the hard-sphere propagator 
derived by Cao and Berne~\cite{CB,KL}. 
Unfortunately, it is impractical to develop an analytic 
pair-product form for the anisotropic dipole-dipole interaction potential.
Thus the primitive approximation, which separates the density matrix into 
kinetic and potential energy terms, was used. In order to choose the value of 
the number of time slices $M$, we performed consistency checks by varying $M$
to see that the results had converged. We used up to 160 time slices 
($M$ = 160) depending on the temperature and the number of particles.
We employed the canonical ensemble, 
${\it i.e.}$ in each simulation we fixed the temperature $T$, 
the number of particles $N$, the diameter of a hard sphere $a_{s}$,
and the strength of the dipolar interactions $\mu$.
The thermal expectation value of an operator $O$ is given by 
$\langle O \rangle = \mbox{Tr} (\rho_B O)/\mbox{Tr} (\rho_B)$. Expectation
values are calculated at the same time the discrete paths are sampled.
Typically 20,000 - 30,000 MC steps are required 
for equilibration and statistical 
averages are collected from 150,000 - 180,000 MC steps after this. 
In each MC step, we attempted 60 - 1,250 trial moves at each time slice.
Multiple runs were performed 
for each set of parameters so that statistical error bars could be determined.
We performed these calculations partly on IBM Power4 and partly on Athlon
processors at the University of Georgia. For $N$ = 125 dipolar particles and 
$\tau$ = 0.00446, 
the simulation (thermalization + computing averages) at a single temperature
took about 2100 IBM Power4 CPU hours.
The density profile $n(R)$, which indicates the likelihood of finding 
particles at a given distance $R$ from the center of the trap, is defined as
\begin{eqnarray}
n(R)  =
\left\langle
\frac{1}{M}\sum_{\kappa=0}^{M-1}\sum_{i=1}^N
\frac{\delta(R_{i,\kappa}-R)}{R^{2}} \right\rangle,
\end{eqnarray}
where $R_{i,\kappa}$ denotes the distance between the position of the $i$th
particle and the center of the trap at time slice $\kappa$ and 
the symbol $<\cdot>$ denotes an ensemble average.
The superfluid fraction along the axis of rotation ${\bf n}$ is given 
by~\cite{PC}
\begin{eqnarray}
\frac{\rho_s}{\rho} = \frac{4m^2}{\beta \hbar^2 I^0}
\langle A_{{\bf n}}^2 \rangle,
\label{sfd}
\end{eqnarray}
where $A_{{\bf n}} = {\bf A}\cdot{\bf n}$ denotes the projected area,
\begin{eqnarray}
{\bf A} = \frac{1}{2}\sum_{\kappa=0}^{M-1} \sum_{i=1}^N
({\bf R}_{i,\kappa} \times {\bf R}_{i,\kappa+1}),
\label{pad}
\end{eqnarray}
and $I^0$ the classical moment of inertia,
\begin{eqnarray}
I^0 = \left \langle m \sum_{\kappa=0}^{M-1} \sum_{i=1}^N \frac{1}{M}
({\bf n} \times {\bf R}_{i,\kappa})\cdot
({\bf n} \times {\bf R}_{i,\kappa+1}) \right \rangle.
\end{eqnarray}

\begin{figure}[]
\begin{picture}(0,380)(0,0)
\put(-130,200){\includegraphics{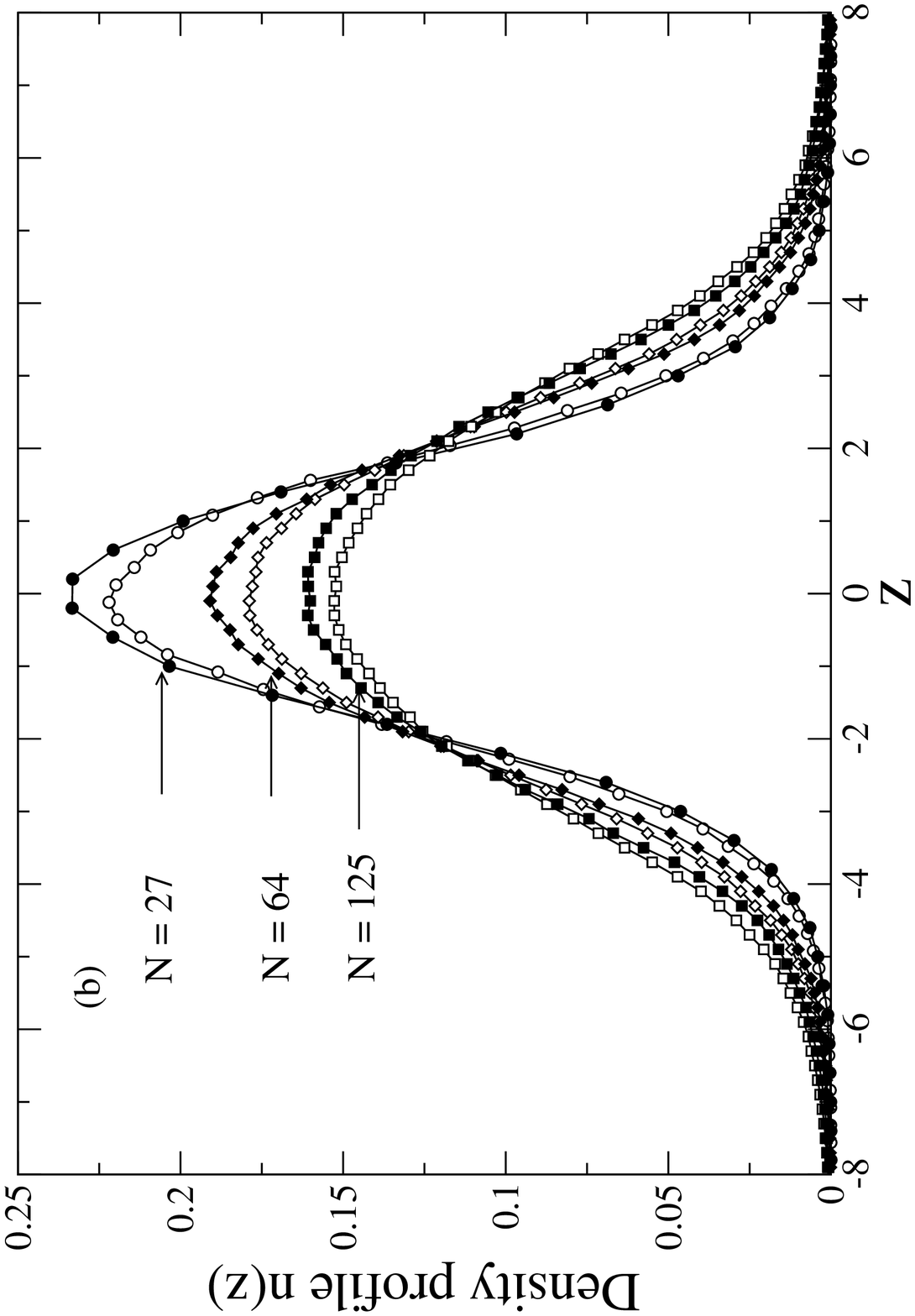}}
\put(-130,390){\includegraphics{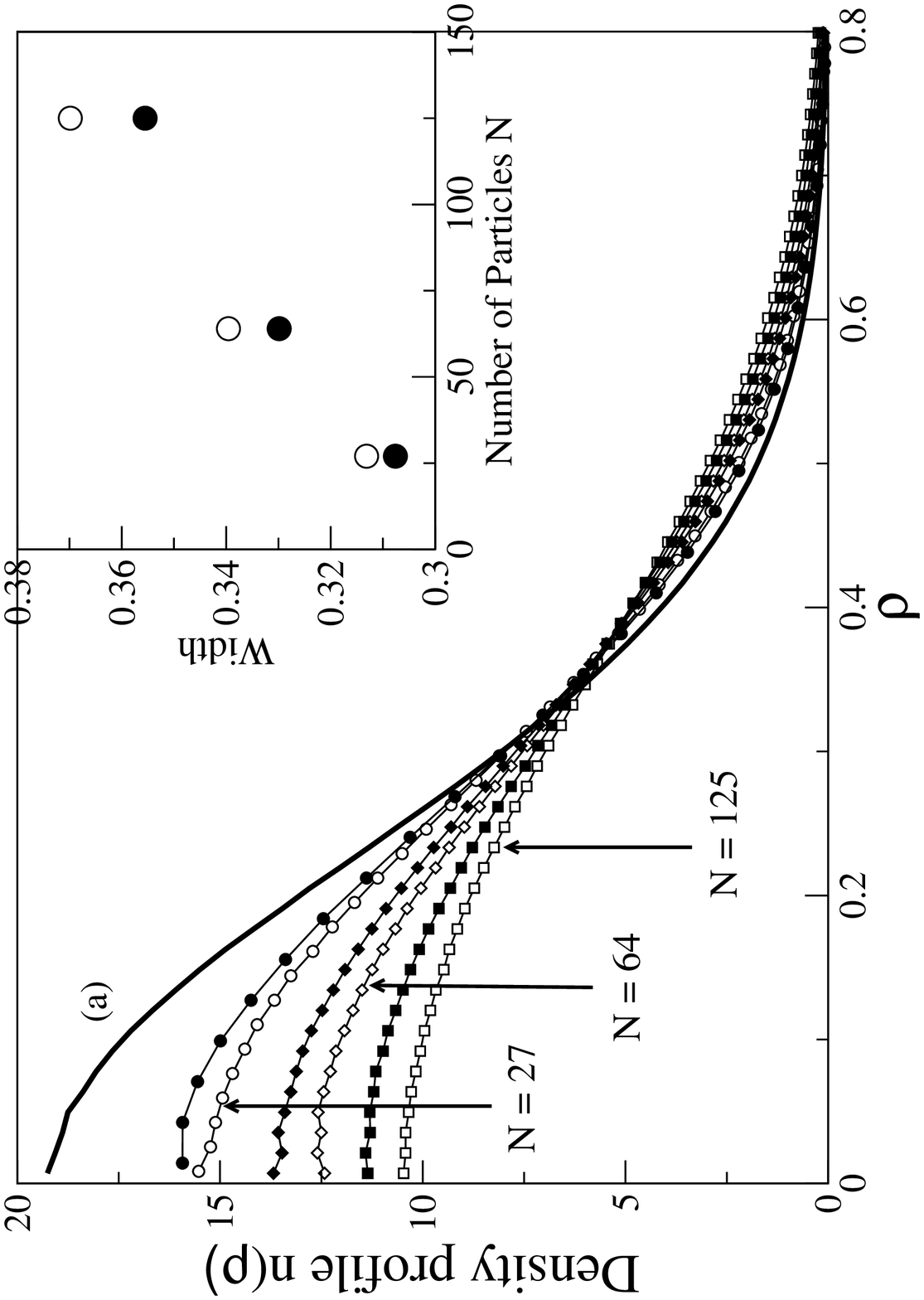}}
\end{picture}
\caption{The calculated density profiles in a cigar-shaped trap 
(a) $n(\rho)$ and (b)$n(z)$ 
at $T$ = 0.4 $T^{3D}_{c}$ for three different numbers of particles, 
$N$ = 27, 64, and 125. The inset in (a) shows half widths.
For comparison, we used two values $\mu$ = 0 (open symbols) 
and $\mu$ = 0.2 (filled symbols). A thick solid line in (a) is the density 
profile $n(\rho)$ for an ideal Bose gas. 
In all figures, when statistical errors cannot be seen 
on the scale of the figure, the error bars are smaller than the symbol sizes.}
\label{fig1}
\end{figure}

\begin{figure}[]
\begin{picture}(0,190)(0,0)
\put(-130,200){\includegraphics{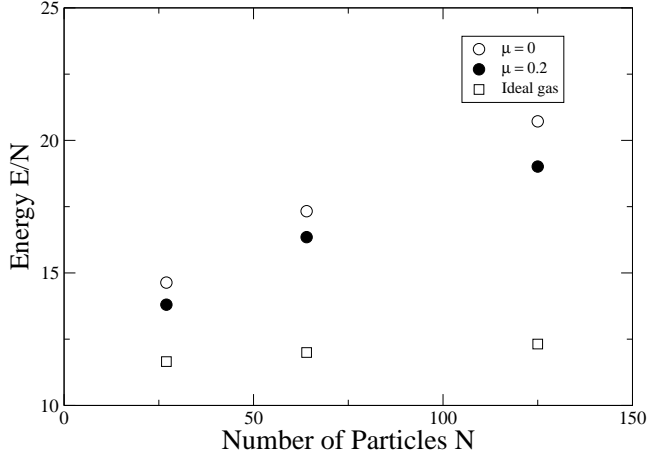}}
\end{picture}
\caption{In a cigar-shaped trap, 
the total energy per particle $E/N$ as a function of the number of 
particles $N$ at $T$ = 0.4 $T^{3D}_{c}$ for $\mu$ = 0 (open circles), 
$\mu$ = 0.2 (filled circles), and an ideal Bose gas (open squares).}
\label{fig2}
\end{figure}

\section{Simulation Results and Discussion}
\label{sec3}
The path integral Monte Carlo (PIMC) method allows one to calculate accurate
quantum mechanical expectation values of many-body systems, the only 
input being the many-body potential. Here, we use PIMC to simulate 
a system composed of dipolar hard spheres for various
values of the number of hard spheres $N$ and the trap aspect ratio $\lambda$
at two fixed dipolar moments $\mu$ as a function of temperature $T$ and
to obtain simulation results for a Bose-Einstein condensation with magnetic
dipole-dipole forces. This section discusses the energetics and structural 
properties of trapped dipolar hard spheres obtained from our PIMC simulations.
For simplicity, throughout this paper, we use a cylindrically 
symmetric harmonic
trap with an aspect ratio $\lambda$, $\lambda = \omega_{x}/\omega_{z}$ and 
$\omega_{x}=\omega_{y}$. The approximate critical temperature $T^{3D}_{c}$ 
for an ideal gas of $N$ atoms in a trap is given by~\cite{DGPS} 
\begin{eqnarray}
1 &=& \left(\frac{T^{3D}_{c}}{T^{3D}_{0}}\right)^{3}
  + \frac{3\bar{\omega}g(2)}{2\omega_{ho}\left[g(3)\right]^{2/3}}
\left(\frac{T^{3D}_{c}}{T^{3D}_{0}}\right)^{2}N^{-1/3}, \\
& & T^{3D}_{0} = \frac{\hbar\omega_{ho}}{k_{B}}\left(\frac{N}{g(3)}\right)^{1/3},\nonumber
\end{eqnarray}
where $T^{3D}_{0}$ is the critical temperature for an ideal Bose gas in a trap
in the thermodynamic limit,
$g(n)$ is the Riemann zeta function, and
$\bar{\omega} = (\omega_{x}+\omega_{y}+\omega_{z})/3 $ and 
$\omega_{ho} = (\omega_{x}\omega_{y}\omega_{z})^{1/3}$ are the arithmetic
average and the geometric average of the trapping frequencies, respectively.
A measure of the strength of the dipolar interaction relative
to the $s$-wave scattering in general is given by 
\begin{equation}
\varepsilon_{dd} = \frac{\mu^{2}m}{3\hbar^{2}a_{s}}.
\end{equation}
For chromium, the more common bosonic isotope, $^{52}$Cr, has 
$\varepsilon_{dd}$ $\approx$ 0.089
and the less common bosonic isotope, $^{50}$Cr, has 
$\varepsilon_{dd}$ $\approx$
0.36~\cite{SHWGGPS}. In this study, in order that the effects of dipolar 
interactions can be made visible,
we used $m$ = $m(^{87}$Rb), 
$\omega_{z}$ ($\omega_{x}$) = 2$\pi \times$ 77.78 Hz, 
$\omega_{x}$ = 10 $\omega_{z}$ ($\omega_{z}$ = 10 $\omega_{x}$),
the $s$-wave scattering length $a_{s}$ = 10 $a_{Rb}$ = 0.0433 $a_{z}$ 
($a_{x}$),
and the dipole moment 
$\mu$ = 0.2 $(\hbar\omega_{z}a_{z}^{3})^{1/2}$ ($(\hbar\omega_{x}a_{x}^{3})^{1/2}$) for a cigar-shaped trap 
(a disk-shaped trap) as an example for numerical calculations, 
giving  $\varepsilon_{dd}$ $\approx$ 0.31. Here, 
$a_{Rb}$ = 100 $a_{0}$ (Bohr radii) is 
the $s$-wave scattering length of $^{87}$Rb,
$a_{z}$ = $\sqrt{\hbar/m\omega_{z}}$, and 
$a_{x}$ = $\sqrt{\hbar/m\omega_{x}}$.
For these parameters, the short-range interactions and the dipole-dipole ones 
have a comparable strength, and we can see the interplay between short-range 
and dipole-dipole interactions and the deviation between the case with and 
without dipole-dipole interactions. The results are scalable.
Moreover, the magnitude and sign
of the values of $a_{s}$ and $\mu$ can be easily tuned 
by Feshbach resonance~\cite{FR} 
and by an external field~\cite{GGP}, respectively.

\begin{figure}[]
\begin{picture}(0,190)(0,0)
\put(-130,200){\includegraphics{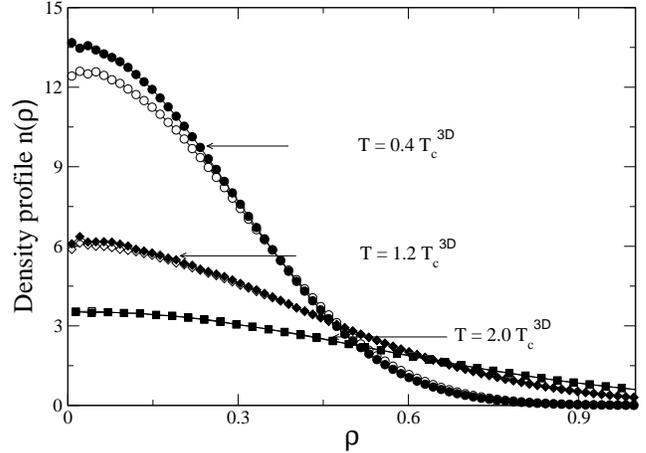}}
\end{picture}
\caption{In a cigar-shaped trap, 
the dependence of the density profile $n(\rho)$ on the temperature $T$
for $N$ = 64.}
\label{fig3}
\end{figure}

\begin{figure}[]
\begin{picture}(0,200)(0,0)
\put(-130,200){\includegraphics{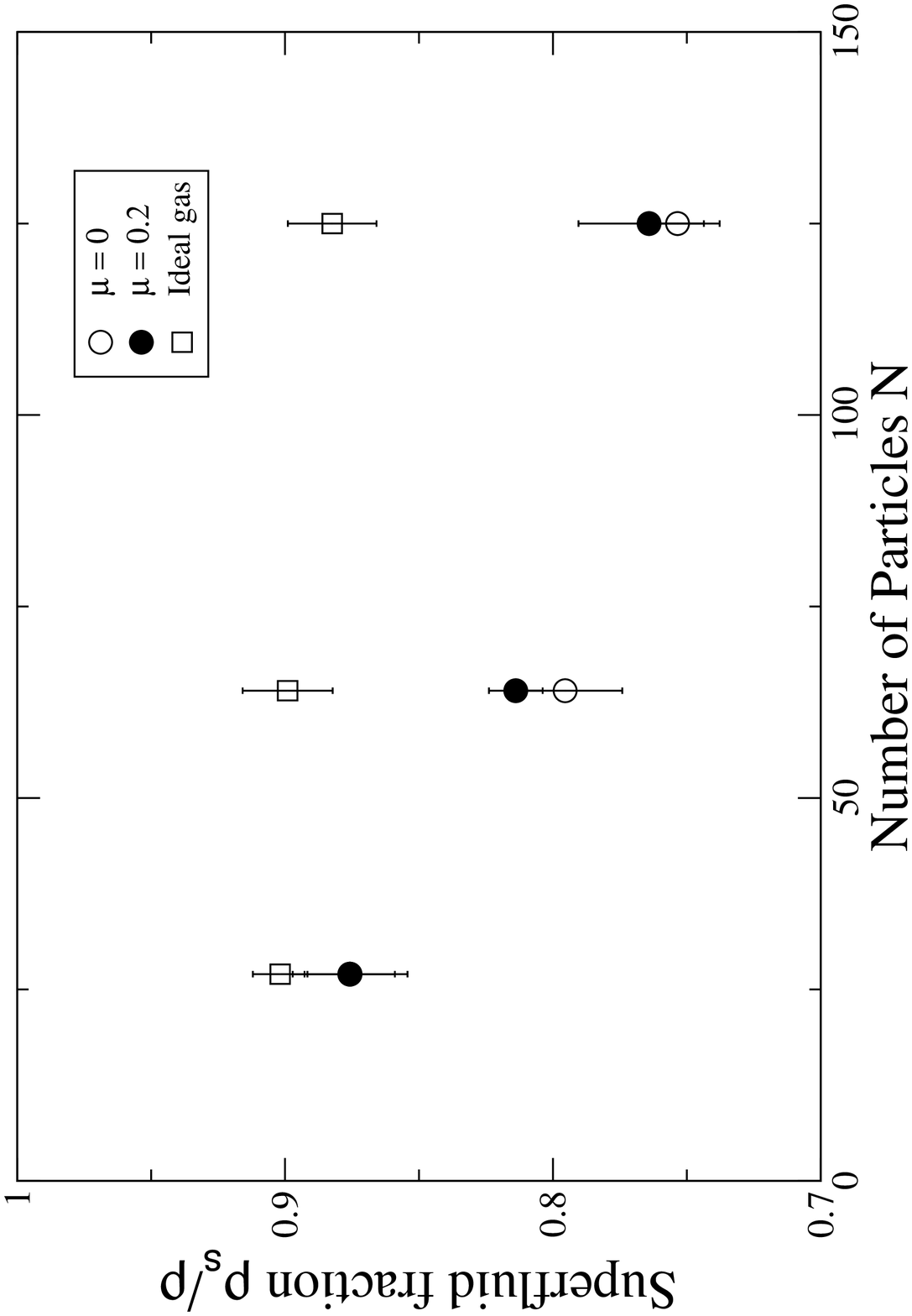}}
\end{picture}
\caption{In a cigar-shaped trap,
the superfluid fraction $\rho_{s}/\rho$ along the axis of rotation
${\bf z}$ as a function of the number of particles $N$ 
at $T$ = 0.4 $T^{3D}_{c}$ for $\mu$ = 0 (open circles), 
$\mu$ = 0.2 (filled circles), and an ideal Bose gas (open squares).}
\label{fig4}
\end{figure}

\begin{figure}[]
\begin{picture}(0,400)(0,0)
\put(-130,225){\includegraphics{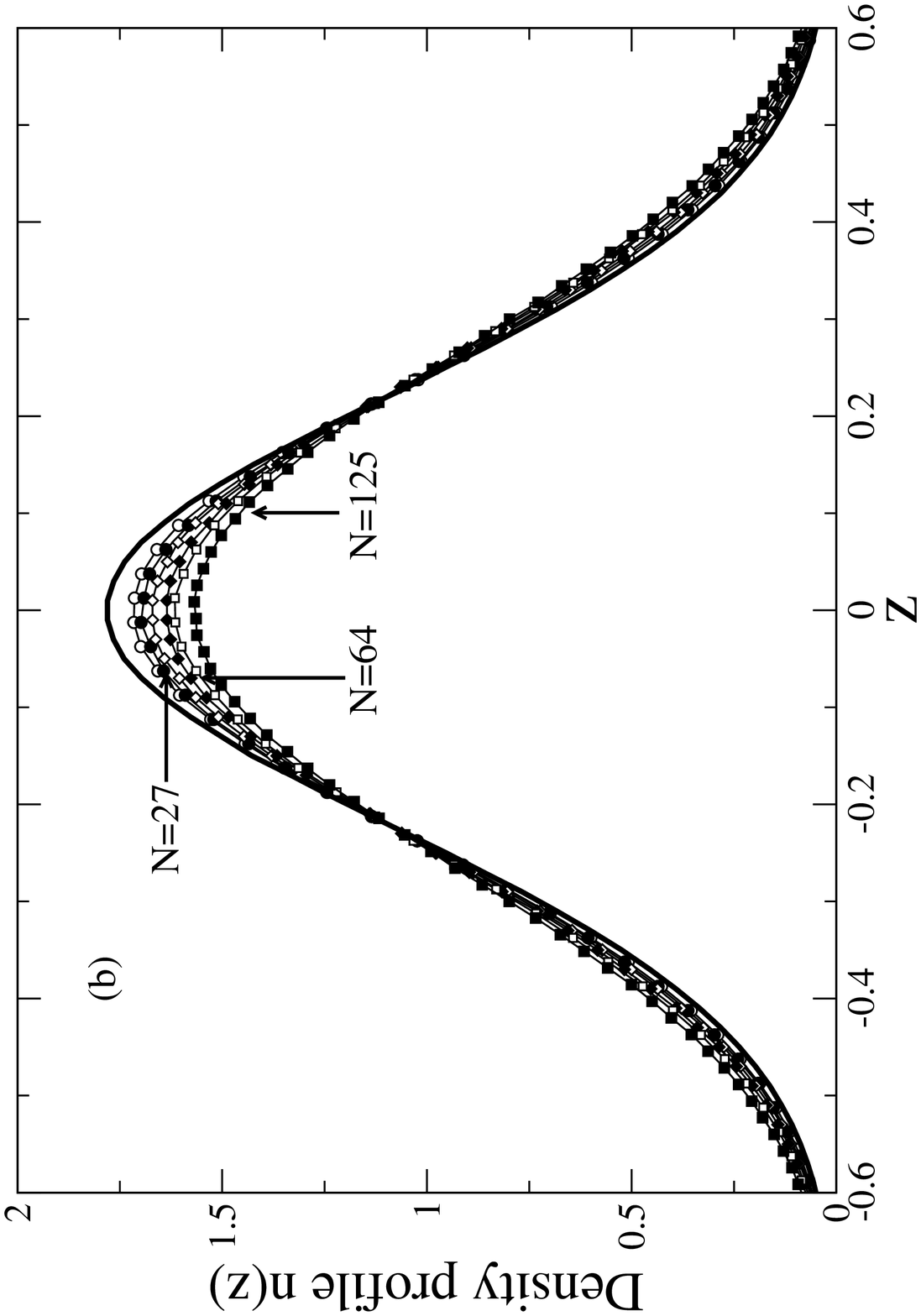}}
\put(-130,410){\includegraphics{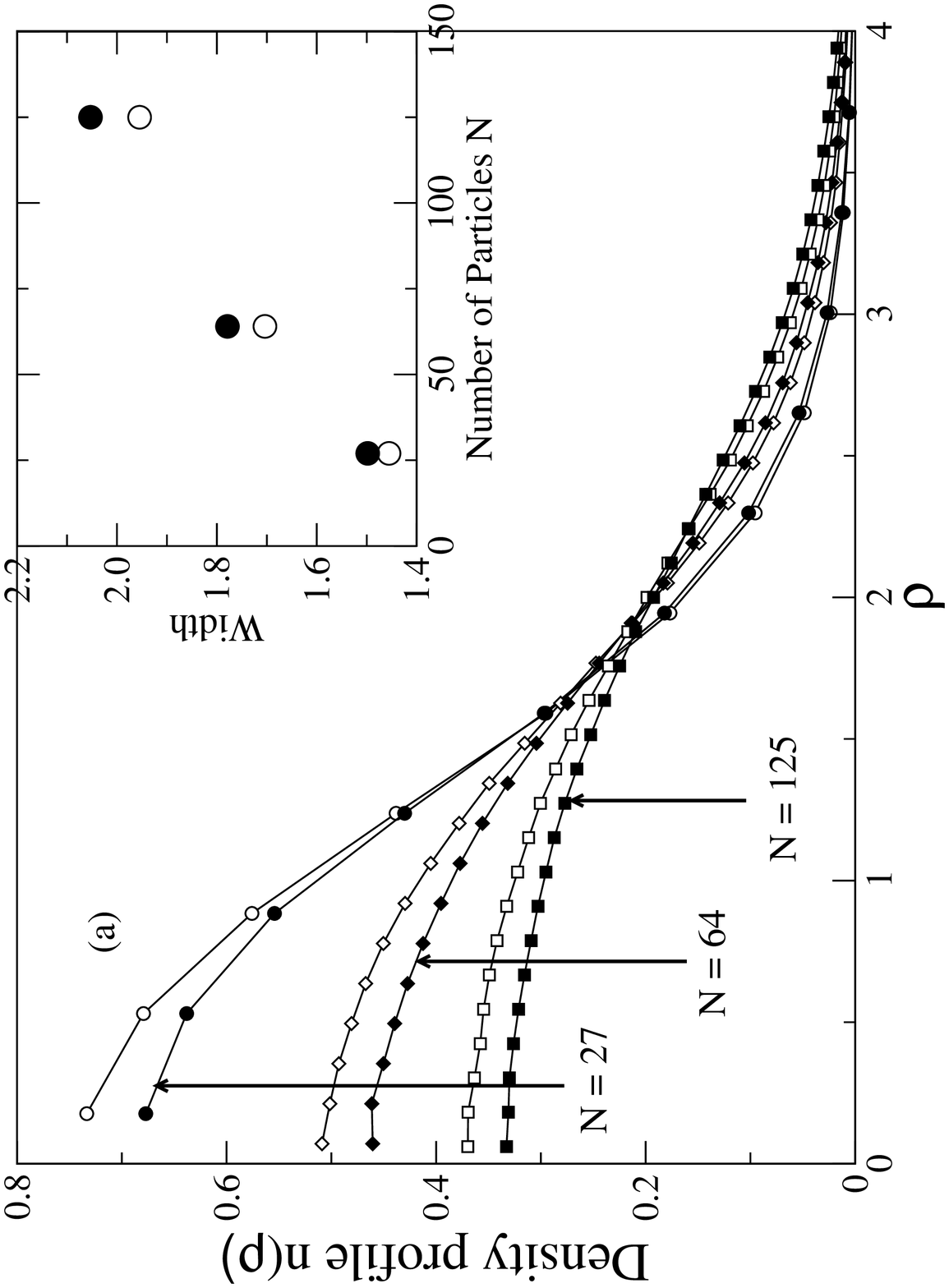}}
\end{picture}
\caption{The calculated density profiles in a disk-shaped trap
(a) $n(\rho)$ and (b) $n(z)$ at $T$ = 0.4 $T^{3D}_{c}$ 
for three different numbers of particles, 
$N$ = 27, 64, and 125. 
For comparison, we used two values $\mu$ = 0 (open symbols) and 
$\mu$ = 0.2 (filled symbols). A thick solid line in (b) is the density 
profile $n(z)$ for an ideal Bose gas.
The inset in (a) shows half widths.}
\label{fig5}
\end{figure}

\begin{figure}[]
\begin{picture}(0,190)(0,0)
\put(-130,200){\includegraphics{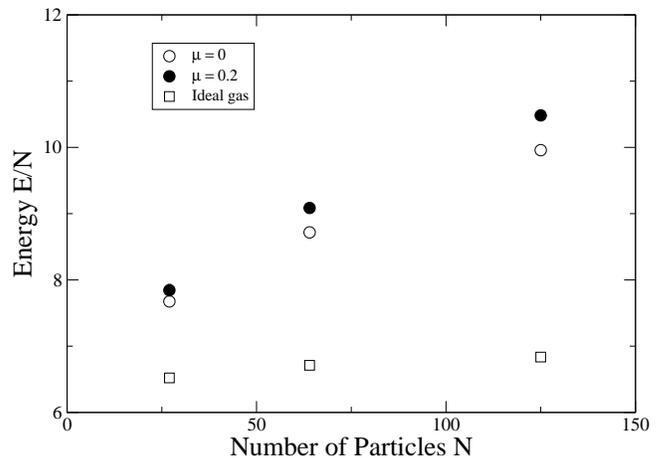}}
\end{picture}
\caption{In a disk-shaped trap, 
the energy per particle $E/N$ as a function of the number of particles $N$
at $T$ = 0.4 $T^{3D}_{c}$ for $\mu$ = 0 (open circles), 
$\mu$ = 0.2 (filled circles), and an ideal Bose gas (open square).}
\label{fig6}
\end{figure}

\subsection{A cigar-shaped trap}
We first consider a 3D dipolar Bose gas in a cigar-shaped trap with 
$\lambda$ = $\omega_{x}/\omega_{z}$ = 10 and $\omega_{x} = \omega_{y}$ 
by increasing the confinement in the transverse direction.
The length unit is $a_{z}$ = $\sqrt{\hbar/m\omega_{z}}$ and 
energies are measured in units of $\hbar \omega_{z}$. 
Figure 1(a) shows the calculated density profiles
$n(\rho)$ as a function of $\rho$,  
normalized such that $\int_{0}^{\infty} n(\rho)\rho d\rho = 1$
and $\rho = \sqrt{x^{2}+y^{2}}$, 
at $T$ = 0.4 $T^{3D}_{c}$ for three different numbers of particles, 
$N$ = 27 (circles), 64 (diamonds), and 125 (squares). 
For comparison, we used two values $\mu$ = 0 and $\mu$ = 0.2.
Filled symbols indicate the density profiles $n(\rho)$ 
for $\mu$ = 0.2 and open symbols the density profiles $n(\rho)$ 
for $\mu$ = 0. 
In all figures, when statistical errors cannot be seen on the scale of 
the figure, the error bars are smaller than the symbol sizes.
Qualitatively, the density profiles for 
$\mu$ = 0 and for $\mu$ = 0.2 are similar in shape, as predicted. However, 
the density profiles $n(\rho)$ for $\mu$ = 0 are broader than 
those for $\mu$ = 0.2. 
The difference with respect to the density profile $n(\rho)$ 
without dipolar interactions is clearly 
seen as the number of particles increases. The density profiles $n(z)$ 
along the axial direction as a function of $z$,
normalized such that $\int_{-\infty}^{\infty} n(z) dz = 1$,
at $T$ = 0.4 $T^{3D}_{c}$ for three different numbers of particles 
are also shown in Fig. 1(b).
In contrast to the density profile $n(\rho)$, 
the difference in the density profiles $n(z)$ 
between for $\mu$ =0 (open symbols)
and for
$\mu$ = 0.2 (filled symbols) is small as the number of particles grows. 
In addition, a thick solid line in Fig. 1(a) shows the density profile 
$n(\rho)$ for an ideal Bose gas for comparison.
The density profile $n(\rho)$ for the interacting gas approaches that
for an ideal gas as the number of particles $N$ decreases. Correspondingly
the excitations in the longitudinal direction are largely frozen out and 
the system behaves quasi-one dimensionally.
Our density profile studies
support the facts that i) as we add more particles into the gas, 
the density profiles expand; the half-width of the density profile
increases with the number of particles $N$, 
as shown in the inset of Fig 1(a) and 
ii) the net effect of dipole-dipole interactions 
in the cigar-shaped trap is attractive, {\it i.e.}
shrinking of the density profiles is observed.
To further characterize the behavior of particles with dipolar
interactions at $T$ = 0.4 $T^{3D}_{c}$, Fig. 2 shows
the total energy per particle $E/N$ of the system 
as a function of the number of particles $N$.
The total energy per particle is strongly affected by the particle-particle 
interaction and the overall dipole-dipole interaction is attractive 
in the cigar-shaped trap, because the total energy per particle for 
$\mu$ = 0.2 is smaller than that for $\mu$ = 0.
From Fig. 2 we can clearly see the dependence on the number 
of particles $N$, and numerical calculations have previously shown 
that the finite-size effects are significant 
for rather small values of $N$~\cite{KL,KND}.
The difference in the total energy per particle $E/N$ between for an ideal 
Bose gas and for an interacting gas increases with increasing $N$,
which is a consequence of the expansion of the density profiles 
with interaction, as shown in Fig. 1,
although the ideal bosonic gas energy per particle $E/N$ changes 
considerably less as the number of particles $N$ increases.
Accordingly, the dependence of the total energy per particle $E/N$ 
on the number of 
particles $N$ reflects essentially the interaction energy alone, 
as the kinetic energy depends weakly on $N$.
Figure 3 shows the dependence of the density profiles $n(\rho)$ 
on the temperature $T$ for $N$ = 64.
As the temperature increases, the effect of interactions is vanishingly small.
At $T$ = 2.0 $T^{3D}_{c}$, the density profiles $n(\rho)$ for $\mu$ = 0 
and for $\mu$ = 0.2 are indistinguishable and the density profiles have 
the Gaussian profiles of a pure thermal distribution.  
Finally, we calculate the superfluid fraction $\rho_{s}/\rho$ along the axis
of rotation ${\bf z}$ and in Fig. 4, 
we plot the superfluid fraction $\rho_{s}/\rho$ as a function of
the number of particles $N$ at $T$ = 0.4 $T^{3D}_{c}$. In addition, 
Fig. 4 shows the superfluid fraction $\rho_{s}/\rho$ for an ideal Bose gas.
In general, 
the superfluid fraction is essentially unity at very low temperatures and 
decreases gradually with increasing temperature.
For the ideal gas, the superfluid fractions show at most a very weak 
dependence on the number of particles considered. 
However, there are significant finite-size effects
for the interacting gas. As the presence of repulsive interactions has 
the effect of lowering the value of the critical temperature, the superfluid
fraction for the interacting gas is smaller than that for the ideal gas 
at the same temperature $T/T^{3D}_{c}$, 
as shown in Fig. 4. The dependence of the superfluid fraction $\rho_{s}/\rho$
on the presence of interactions increases with 
the number of particles $N$.
It is not easy, however, 
to see the difference between the superfluid fractions 
$\rho_{s}/\rho$ for $\mu$ = 0.2 (filled circles) 
and those for $\mu$ = 0 (open circles) because of the size of the errorbars
but the difference must clearly be small, if any. 

\begin{figure}[]
\begin{picture}(0,175)(0,0)
\put(-130,200){\includegraphics{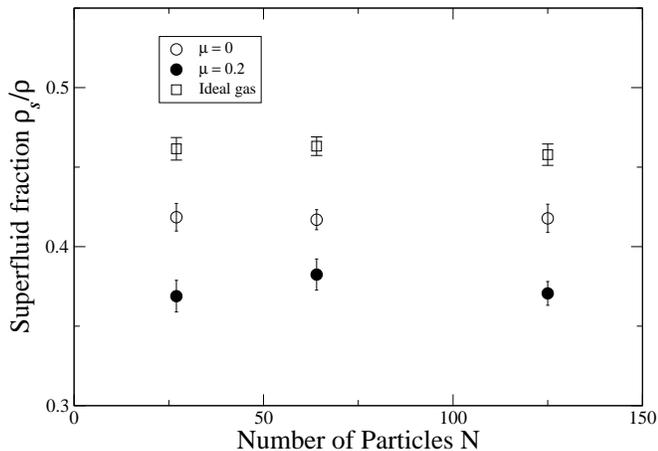}}
\end{picture}
\caption{In a disk-shaped trap, the superfluid fraction 
$\rho_{s}/\rho$ along the axis of rotation
${\bf z}$ as a function of the number of particles $N$ 
at $T$ = 0.4 $T^{3D}_{c}$ for $\mu$ = 0 (open circles), 
$\mu$ = 0.2 (filled circles), and an ideal Bose gas (open squares).}
\label{fig8}
\end{figure}

\subsection{A disk-shaped trap}

For the disk-shaped trap with the asymmetry factor $\lambda$ = 
$\omega_{x}/\omega_{z}$ = 0.1, {\it i.e.} increasing the confinement 
in the axial direction, 
the length unit is $a_{x}$ = $\sqrt{\hbar/m\omega_{x}}$ and the energy unit
$\hbar\omega_{x}$. 
In Fig. 5, as we did for the cigar geometry,
we plot the density profiles $n(\rho)$ and $n(z)$ as a function of $\rho$ and 
$z$, respectively, at $T$ = 0.4 $T^{3D}_{c}$ 
for three different numbers of particles, $N$ = 27 (circles), 64 (diamonds), 
and 125 (squares) using $\mu$ = 0 (open symbols) 
and $\mu$ = 0.2 (filled symbols).
It is compared to corresponding density profiles $n(\rho)$ and $n(z)$ for
a Bose gas with short-range interactions only and for an ideal bosonic gas.
Compared to the density profiles for $\lambda$ = 10, they are similar 
in shape. As $\omega_{z}$ is larger than $\omega_{x}$, 
the density profile $n(z)$ for an interacting gas approaches 
that for an ideal gas (see Fig 5(b)), 
indicating the excitations 
in the axial direction are largely frozen out and the system behaves 
quasi-two dimensionally.
In a disk-shaped trap, 
the density profiles expand along both the $z$-axis and the $\rho$-axis
as we increase $\mu$, in agreement with those found by other authors.
Correspondingly, the net contribution of long-range dipolar forces is 
repulsive, in contrast to the cigar-shaped trap.
Simulations for other values of $\lambda$ show that
BEC in a disk-shaped trap is stable, even in very asymmetric traps,
because the dipole-dipole interaction energy always
remains positive.
Figure 6 shows the total energy per particle $E/N$
as a function of the number of particles $N$ both for interacting atoms with
$\mu$ = 0 (open circles) and $\mu$ = 0.2 (filled circles) 
and for noninteracting
particles (open squares).
The total energy per particle $E/N$ for $\mu$ = 0.2 is larger than 
that for $\mu$ = 0 due to the increasing repulsive interactions.
The total energy per particle $E/N$ for an interacting Bose gas increases 
more rapidly than that for an ideal bosonic gas 
due to the finite-size effect, as $N$ increases.
Therefore, in the Thomas-Fermi limit the kinetic energy of the particles 
in the trap is negligible in comparison to the interparticle interaction 
energy and the trapping potential.
In addition, Fig. 7 shows the superfluid fraction
$\rho_{s}/\rho$ along the axis of rotation ${\bf z}$ as a function
of the number of particles $N$ at $T$ = 0.4 $T^{3D}_{c}$. 
The superfluid fraction $\rho_{s}/\rho$ for an noninteracting 
gas is also shown in the figure. 
The superfluid fraction $\rho_{s}/\rho$ 
for $\mu$ = 0.2 (filled circles) is clearly smaller than 
that for $\mu$ = 0 (open circles). 
The superfluid fraction $\rho_{s}/\rho$ shows no dependence on the number of 
particles $N$, whereas the interparticle interaction plays a significant role 
in decreasing of the superfluid fractions $\rho_{s}/\rho$.
Interestingly, when we compare the superfluid fraction 
for $\lambda$ = 0.1 to that for $\lambda$ = 10, we find that 
the superfluid fraction in the disk-shaped trap is significantly smaller than
that in the cigar-shaped trap because geometric arguments imply 
that more excited longitudinal modes are 
occupied for $\lambda$ = 0.1 than for $\lambda$ = 10.

\section{Conclusion}
\label{sec4}

In this paper, Bose-Einstein condensation of trapped dipolar Bose gases 
has been studied using a finite-temperature path integral Monte Carlo 
technique. The quantum particles have a dipole-dipole interaction and 
a $s$-wave scattering interaction. The dipolar potential energy is long-range,
anisotropic, and partially attractive, compared to the short-range and purely
repulsive $s$-wave scattering interaction.
Using pseudopotential forms to describe the interatomic interactions, 
we have calculated the equilibrium properties, 
such as the energetics and structural properties,
of a system composed of $N$ dipolar hard spheres. 
We find that 
for a cigar-shaped trap, the net effect of dipolar forces is attractive and 
shrinking of the density profiles is observed; and for a disk-shaped trap, 
the net contribution of dipole-dipole interactions is repulsive and the density
profiles expand. Accordingly, the net contribution of dipolar interactions 
depends on the trapping aspect ratio $\lambda$.

\begin{acknowledgments}
We are greatly indebted to P. Stancil for his critical reading of 
the manuscript and profound comments. This work was partially supported 
by NASA grant No. NNC04GB24G.

\end{acknowledgments}

\end{document}